\documentclass[12pt,preprint]{aastex}

\begin{document}

\title {POLLUTION OF SINGLE WHITE DWARFS BY ACCRETION OF MANY SMALL ASTEROIDS}

\author{M. Jura \\ Department of Physics and Astronomy and Center for Astrobiology  \\ University of California, 
 Los Angeles CA 90095-1562; jura@astro.ucla.edu}
\begin{abstract}
Extrapolating from  the solar system's asteroid belt, we propose that  externally-contaminated white dwarfs without an infrared excess may be experiencing continuous accretion of gas-phase   material that   ultimately is derived from the tidal destruction of multiple small asteroids.  If this scenario is correct, then observations  of metal-polluted white dwarfs  may 
lead to determining  the bulk elemental compositions of ensembles of extrasolar minor planets.
  \end{abstract}
 \section{INTRODUCTION}
 An indirect but potentially powerful method to determine the elemental abundances of extrasolar minor bodies and  planets is to measure
 the photospheric compositions of stars  they pollute.   Such  studies  in main-sequence stars are  challenging because
 the host star's photosphere  possesses  intrinsic metals, making it difficult to determine the amount, if any, of contamination (see Gonzalez 2006).  In contrast,  heavy elements rapidly  settle out of the atmospheres of white dwarfs cooler than 20,000 K (Koester \& Wilken 2006), so that photospheric metals have an external source.  We hope to identify which  metal-possessing cool white dwarfs have accreted circumstellar rather than interstellar material in order to use these white dwarf atmospheres to measure the bulk elemental  composition of extrasolar 
minor planets.

There
 are strong arguments that  the approximately ten single white dwarfs known to have an infrared excess  (Kilic et al. 2006a, 2006b, Kilic \& Redfield 2007, Mullally et al. 2007, Jura et al. 2007b, von Hippel et al. 2007, Farihi et al. 2008a, b) are accreting tidally-disrupted asteroids.   (1)  The near-infrared excess discovered around G29-38 (Zuckerman \& Becklin 1987) was shown to be produced by dust (Graham et al. 1990) that likely resulted from  an orbitally-perturbed minor planet (Debes \& Sigurdsson 2002) that passed within the tidal-radius of the host star (Jura 2003). For other white dwarfs, the infrared excess is usually well modeled with a geomerically flat dust disk that lies within the tidal-radius of the star (Jura  et al. 2007b).  (2)   The infrared spectrum of G29-38 displays strong 10 ${\mu}$m silicate emission with a spectral shape that resembles dust emission from the zodiacal light rather 
 than interstellar  silicates (Reach et al. 2005).  GD 362, the second white dwarf discovered to have a near-infrared excess (Becklin et al. 2005, Kilic et al. 2005), also displays  a 10 ${\mu}$m silicate
 feature characterisitic of asteroidal rather than interstellar dust (Jura et al. 2007a).  (3)  The white dwarfs with an infrared excess generally show when measured that F$_{\nu}$(24 ${\mu}$m) $<<$ F$_{\nu}$(8 ${\mu}$m) (Reach et al. 2005, Jura et al. 2007b).  There is no evidence for relatively cold dust grains outside the tidal-radius (Jura et al. 2007b).   (4) The atmospheric
 composition of GD 362 is refractory-rich and volatile-poor (Zuckerman et al. 2007), reminiscent of material in the inner solar system and unlike the interstellar medium.  While not as comprehensively studied as GD 362, there are at least three other white dwarfs which are makedly deficient in carbon relative to iron as is characteristic of asteroids and very unlike the Sun or interstellar matter (Jura 2006).
 (5)  Asteroid belts similar  the solar system's can supply the required pollution masses (Jura 2006).

 Only about 1/7 of all contaminated white dwarfs display an infrared excess (Kilic \& Redfield 2007);  the origin
 of the pollution of the  more numerous contaminated white dwarfs without an infrared excess is  uncertain.  Although interstellar accretion could supply the metals in the atmospheres of contaminated white dwarfs
(see, for example, Dupuis et al. 1993, Koester \& Wilken 2006, Zuckerman et al. 2003, Dufour et al. 2007), there are difficulties with this model.  First, at least for  helium-rich stars, the accretion of hydrogen relative to calcium must be suppressed by at least a factor of 100  (Dufour et al. 2007).  None of the suggested mechanisms such as magnetic shielding that preferentially exclude hydrogen
from the star have yet received any observational support (Friedrich, Jordan \& Koester 2004).  Furthermore, there is no correlation between metal contamination and the star's space motion as would be expected from Bondi-Hoyle accretion theory (Zuckerman et al. 2003, Koester \& Wilken 2006).  Finally,  using a very
simple model of grain inspiral because of Poynting-Robertson drag, Jura et al. (2007b) showed
that the predicted fluxes at 24 ${\mu}$m for Bondi-Hoyle accretion can be more than 100 times greater than
the observed upper limits.  In view of  these difficulties, the alternative of accretion of
circumstellar matter merits consideration.

Gaensicke et al. (2006, 2007) have identified two  metal-contaminated single white dwarfs with gaseous dust disks that are likely the result of tidal-disruption of a large parent-body such as an asteroid.
It is possible that in many cases, the dust  orbiting a white dwarf is destroyed and therefore stars
without an infrared excess may still experience accretion from tidally-disrupted minor bodies.  In this paper, we elaborate upon this possibility.  

In contrast to hydrogen-rich white dwarfs where the gravitational settling time of metals is typically less than 100 years (see von Hippel \& Thompson 2007), in helium-rich white dwarfs cooler than 20,000 K, the gravitational settling
time may be considerably longer than 10$^{5}$ yr (Paquette et al. 1986, Dupuis et al. 1993).  Although very uncertain, below, we suggest that the lifetime of a dusty disk around a white dwarf might be
${\sim}$10$^{5}$ yr.  Therefore, at least for the helium-rich white dwarfs, the absence of an infrared
excess might simply be a situation where a single tidally-disrupted asteroid has been fully accreted and
the disk has dissipated but the contamination still lingers in the star's atmosphere.

In ${\S2}$ we present a qualitative  overview of our proposed model where we suggest that contaminated white dwarfs without an infrared excess accrete
from a gaseous disk created by the accumulation from many small asteroids.  This contrasts with our model for the white dwarfs with an infrared excess which we expect to be the result of the destruction of a single large asteroid.  In ${\S3}$ we argue that asteroids larger than 1-10 km in radius are likely to survive the star's evolution through the Asymptotic Giant Branch (AGB) phase.    
  In ${\S4}$, we describe in somewhat more detail our  model of white dwarf contamination by  multiple  tidal-disruptions of  small asteroids.     In ${\S5}$ we discuss observational asssessment  of this scenario.  In ${\S6}$ we discuss implications of this model and in ${\S7}$ we present our conclusions.
 
\section{OVERVIEW}
Very little is known about extrasolar asteroid belts.  Even the presence of warm dust at a few AU from the
star does not tightly constrain the parent body populations (see Wyatt et  al. 2007).  Here, assuming
that many other main-sequence stars possess an asteroid belt similar to the solar system's, we
present a single illustrative model to account for the white dwarf data.  There are many unknown parameters;
 our calculations are aimed to establish there is at least one plausible scenario where multiple 
 small asteroids can explain the presence of metals in  white dwarf photospheres.    If detailed abundance studies
 of polluted white dwarfs without an infrared excess are shown to be consistent with this hypothesis, then a more comprehensive investigation
 would be warranted.  

In our illustrative model, we suggest that the total mass of the asteroid
belt when the star becomes a white dwarf is 10$^{25}$ g, somewhat larger than the solar system's
asteroid belt current mass of 1.8 ${\times}$ 10$^{24}$ g (Binzel, Hanner \& Steel 2000).   We assume there is a steeply rising number of lower-mass asteroids, and 
 we then compute that asteroids
 larger than ${\sim}$1-10 km in radius plausibly survive the star's evolution through the asymptotic giant
branch.   Although we do not compute detailed models, we propose  that the orbits of these asteroids are occasionally perturbed on a time scale of ${\sim}$1 Gyr into orbits
which lead them within the tidal radius of the white dwarf (see Debes \& Sigurdsson 2002, Duncan \& Lissauer 1998) where the asteroid
is rapidly shredded into smaller rocks and dust debris.  

Small asteroids are sufficiently common, that perhaps one is perturbed into an orbit where it is tidally-disrupted every 300 years.  We argue that the accretion time on the white dwarf is long compared to this
arrival time, so that white dwarfs with surviving asteroid belts inevitably possess a low-mass disk composed of disintegrated larger bodies.   It is likely  that newly perturbed asteroids arrive at a  random nonzero inclination
relative to the pre-existing disk.  As the asteroid passes through the disk, solid material is rapidly sputtered into the gas-phase and a  small asteroid is effectively vaporized.  The matter gained by the disk from disrupted asteroids is lost by accretion into the atmosphere of the white dwarf, thus explaining the  star's external pollution.  In these systems, there is  little dust and a negligible infrared excess.  

More rarely, a large asteroid is perturbed into an orbit where it is tidally disrupted.  If sufficiently large,
the destroyed asteroid dominates over the pre-existing gaseous disk.  In these systems, a dust
ring is formed and survives for perhaps 10$^{5}$ years; during this phase there is an infrared excess.

\section{ASTEROID SURVIVAL DURING THE STAR'S AGB PHASE}

Asteroids must survive the AGB phase if they are to pollute the star as a white dwarf.   Rybicki \& Denis (2001) discuss three processes that can lead to terrestrial planet destruction during the AGB phase.  A planet can induce a tidal bulge in its host star which can then interact with the planet to drag it inwards. However, asteroids have such small masses, that this effect is not important.  Below, we consider two other routes
for asteroid destruction.

\subsection{Asteroid Destruction by Wind Drag}

Asteroids may spiral into the host star under the action of gas drag as they orbit through the wind of the AGB star.  We assume
that an asteroid is destroyed if it encounters its own mass in the wind.   This destruction occurs because of
hydrodynamic friction rather than sputtering which is negligible since the colliding gas atoms have kinetic energies of only a few eV (see Tielens et al. 1994).   At distance $D$ from the star in a spherically symmetric wind of mass loss rate ${\dot M_{*}}$ with outflow speed,
$V_{wind}$, the gas density, ${\rho}_{wind}$, is
\begin{equation}
{\rho}_{wind}\;=\;\frac{{\dot M_{*}}}{4{\pi}D^{2}\,V_{wind}}
\end{equation}
As long as gravitational focusing is negligible, the rate of encountering mass, ${\dot M_{enc}}$, by the asteroid  is:
\begin{equation}
{\dot M_{enc}}\;=\;{\pi}\,R_{ast}^{2}\,{\rho}_{wind}\,V_{ast}
\end{equation}
where the asteroid of radius $R_{ast}$ has orbital speed $V_{ast}$.  For this evaluation, we assume that the asteroid's
radius and mass are constant during the AGB phase.  
For circular motion:
\begin{equation}
V_{ast}\;=\;\left(\frac{G\,M_{*}}{D}\right)^{1/2}
\end{equation}

If the time scale for mass loss is  slow compared to the asteroid's orbital period, then the specific angular momentum of the asteroid is approximately conserved and:
\begin{equation}
D\,M_{*}\;=\;D_{i}\,M_{i}
\end{equation}
for an initial distance, $D_{i}$,  from a star of initial mass, $M_{i}$.  As the star loses mass and $M_{*}$ diminishes, then the asteroid's orbit expands.  
Combining equations (1)-(4), then
during the star's AGB phase, the total encountered mass, $M_{enc}$ is:
\begin{equation}
M_{enc}\;=\;{\int}{\dot M_{enc}}\,dt\;{\approx}\;\frac{R_{ast}^{2}\,G^{1/2}\,M_{i}^{3/2}}{16\,V_{wind}\,D_{i}^{5/2}}
\end{equation}
In arriving at the last expression in equation (5), a simplified and slightly altered version of equation (3)  of Duncan \& Lissauer (1998), we assume $M_{i}$ $>>$ $M_{f}$ where $M_{f}$ is the final of the star (see, for example, Weidemann 2000). 

If ${\rho}_{ast}$ is the density of the asteroid whose mass is $(4{\pi}{\rho}_{ast}R^{3}_{ast})/3$,   then equation (5) leads to the condition that the minimum radius of a surviving asteroid, $R_{min}$, is:
\begin{equation}
R_{min}\;=\;\frac{3\,M_{i}^{3/2}\,G^{1/2}}{64\,{\pi}\,D_{i}^{5/2}\,{\rho}_{ast}\,V_{wind}}
\end{equation}

We only consider asteroids with initial distances from the host star greater than 2 AU because interior to this region,  there may be a  quasi-static region around the AGB star with a density  much greater than given by equation (1) (Reid \& Menten 1997).   Beyond 2AU, we assume an outflow velocity of 2 km s$^{-1}$ (Keady, Hall \& Ridgway 1988), a speed that is   appreciably smaller
than the typical wind terminal speed of 15 km s$^{-1}$ (Zuckerman 1980).
We adopt ${\rho}_{ast}$ = 2.1 g cm$^{-3}$, as inferred for Ceres (Michalak 2000). 

 We show in Figure 1 the results for $R_{min}$ as a function of $D_{i}$ for initial stellar masses  of 1.5 M$_{\odot}$, 3 M$_{\odot}$ and 5 M$_{\odot}$.   As can be seen from equation (6),  the value of $R_{min}$ diminishes rapidly as $D_{i}$ increases since less mass is encountered in the wind.  Also, $R_{min}$ increases with the intial mass of the star since the star ejects more mass for the asteroid to encounter. Depending upon the initial distance from the host star and its mass,  we find that  asteroids larger than 1-10 km in radius are likely to survive
the drag induced by the AGB wind.

\subsection{Asteroid destruction by thermal sublimation}

Subjected to the star's high AGB luminosity, an asteroid might thermally sublimate, and we now estimate the minimum size asteroid that survives this process.  If the star's high luminosity phase has  duration, $t_{AGB}$, then, the minimum size of a surviving asteroid is:
\begin{equation}
R_{min}\;=\;\frac{dR_{ast}}{dt}\,t_{AGB}
\end{equation}
$dR_{ast}/dt$ is the rate at which an asteroid's radius shrinks because of sublimation:
\begin{equation}
\frac{dR_{ast}}{dt}\;=\;\frac{{\dot{\sigma}}(T)}{{\rho}_{ast}}
\end{equation}
where  ${\dot {\sigma}}(T)$ denotes the
 mass production rate from the dust per unit area (g cm$^{-2}$ s$^{-1}$) as a function of the asteroid
 temperature, $T$.   Assuming pure olivine, then:
 \begin{equation}
{\dot {\sigma}}(T)\;=\;{\dot {\sigma}}_{0}\,{\sqrt{\frac{T_{0}}{T}}}\,e^{-T_{0}/T}
\end{equation}
where, converted to cgs units,  ${\dot {\sigma}}_{0}$ = 1.5 ${\times}$ 10$^{9}$ g cm$^{-2}$ s$^{-1}$ and $T_{0}$ = 65,300 K  (Kimura et al. 2002).  Different materials may have different sublimation rates so our results are sensitive to the presumed asteroid composition.  We adopt olivine as this appears to be the most likely carrier of the 10 ${\mu}$m silicate emission seen in G29-38 (Reach et al. 2005) and GD 362 (Jura et al. 2007a) as well as being common in the solar system.   
Assuming that the asteroids have a negligible albedo, we write  that:
\begin{equation}
T\;=\;\left(\frac{L_{*}}{16\,{\pi}\,{\sigma}_{SB}\,D^{2}}\right)^{1/4}
\end{equation}
where $L$ is the luminosity of the star and ${\sigma}_{SB}$ is the Stephan-Boltzmann constant.  
While on the AGB, we assume the star has a luminosity near 10$^{4}$ L$_{\odot}$ or larger for  $t_{AGB}$ = 2 ${\times}$ 10$^{5}$ yr (Jura \& Kleinmann 1992).  
     
    Using equation (8)-(10),   we show in Figure 2 the minimum size asteroid from equation (7)  for stellar luminosities of 1 ${\times}$ 10$^{4}$ L$_{\odot}$ and 2 ${\times}$ 10$^{4}$ L$_{\odot}$ as a function of the asteroid's distance from the star.  We neglect the orbital drift of the asteroid discussed above.   Because the thermal sublimation rate is temperature sensitive, asteroid survival decreases rapidly
      as the star's luminosity increases or for asteroids that orbit relatively close to the star.    We see from Figure 2 that depending upon the luminosity of the star, asteroids with radii between 1 and 10 km are likely to survive if they orbit beyond 3 or 4 AU.  
     
\section{ACCRETION OF MANY SMALL ASTEROIDS}

 Eventually, those  asteroids that survive a star's AGB evolution, may have their orbits perturbed so that they come
within the tidal radius of the white dwarf (Duncan \& Lissauer 1998, Debes \& Sigurdsson 2002).  When this occurs, 
 the asteroid is shredded into dust. If the asteroid is relatively small and
if there is a pre-existing disk from destruction of previous asteroids,  then the shredded debris of
the new asteroid is destroyed by sputtering  in the pre-existing disk, and the circumstellar matter is largely gaseous. If, instead, the asteroid is more massive than the pre-existing disk, there is not enough gas to sputter the debris from the
asteroid, and a disk with a large amount of dust at the inclination angle of the large asteroid is created.  This more massive disk subsumes the low-mass pre-existing gaseous disk.  Eventually, however, this
large disk is accreted onto the star and the system relaxes to its steady state configuration of a largely
gaseous disk.

Denote the total asteroid belt  total mass as $M_{belt}(t)$, where $t$ denotes the cooling time of the white dwarf.  Let $t_{orbit}$ be the mean lifetime of
an asteroid before it is perturbed into the tidal radius of the star.  Assuming that all of the shredded asteroid
ultimately is accreted, then the average metal-accretion rate,
$<{\dot M_{metal}}>$, is:
\begin{equation}
<{\dot M_{metal}}>\;=\;\frac{M_{belt}(t)}{t_{orbit}}
\end{equation}

If the circumstellar gaseous disk around a white dwarf has a lifetime, $t_{gas}$, then 
 the mass of a typical disk, $M_{disk}$, is:
\begin{equation}
M_{disk}\;=\;<{\dot M_{metal}}>\,t_{gas}
\end{equation}
  We propose that if an asteroid has a mass larger than $M_{disk}$, then a
single event dominates the circumstellar environment, and the star exhibits an infrared excess.  
If, however, an asteroid arrives with a mass lower than $M_{disk}$, then this newly-shredded asteroid's dust debris is  rapidly vaporized and the disk remains gaseous.  

We picture that all polluted white dwarfs possess in a steady state a gaseous disk, and, occasionally,
a massive dusty disk.  It is likely that a white dwarf takes relatively little time to enter its
``normal" steady state with a gaseous disk.   The first tidally-disrupted asteroid is shredded into dust which can persist for a long time since there is not likely to be much viscosity in the dust particles.
The second tidally-disrupted asteroid also is shredded into dust, but its orbital inclination is likely to
be different from that of the first asteroid. Consequently, there will be grain-grain collisions
approaching ~1000 km s$^{-1}$ within the tidal radius. These collisions are very destructive and large amounts
of gas will be released. Eventually, as described below, the disk will evolve  into a largely gaseous system.

\subsection{Duration of Gaseous Disks: $t_{gas}$}

To progress further, we now estimate the lifetime of a white dwarf's circumstellar disk which we presume is controlled by viscous dissipation.  We could  extrapolate from cataclysmic variables to  single white dwarfs.
However,  disks derive from destroyed asteroids probably have little hydrogen, and their compositions are very different from that of the disks in  cataclysmic
variables.  
For a gaseous circumstellar disk, the time for viscous dissipation is (King, Pringle \& Livio 2007) 
\begin{equation}
t_{gas}\;{\sim}\;D\,\frac{V_{ast}}{{\alpha}V_{th}^{2}}
\end{equation}
where ${\alpha}$ is the usual dimensionless parameter, and  $V_{th}$ is ${\sim}k_{B}T/m_{ast}$ where $k_{B}$ is the Boltzmann constant, $T$ the temperature and $m_{ast}$ the mean atomic weight of the asteroidal matter.    The most uncertain factor is ${\alpha}$.  As discussed by King et al. (2007), empirical estimates of ${\alpha}$ for thin, fully ionized disks are between 0.1 and 0.4, but
theoretical simulations yield values of ${\alpha}$ that are at least an order of magnitude lower.  In FU Ori stars, values of ${\alpha}$ as low as 0.001 are inferred.  The viscosity also may be a function of the gas ionization and decrease as the gas becomes more neutral. Here, because the disks around single white dwarfs may be long-lived, we assume that ${\alpha}$ is 0.001, but this value  is extremely uncertain.  We use models of the infrared emission from disks (Jura et al. 2007b) 
 to adopt a characteristic disk size,  $D$, of 3 ${\times}$ 10$^{10}$ cm.  At this distance from the star,  $V_{ast}$ = 5 ${\times}$ 10$^{7}$ cm s$^{-1}$, and at least for dusty disks, a representative temperature is 300 K  (Jura 2003).  If $m_{ast}$ equals that of silicon, then $t_{gas}$ ${\approx}$ 5 ${\times}$ 10$^{4}$ yr.  
 Although this estimate of the time scale is very uncertain, it is long compared to the time between arrivals of small asteroids within the tidal zone. 

\subsection{Destruction timescale of Small Asteroids in a Gaseous Disk}

We now show that if a relatively small asteroid impacts a gaseous disk that the resulting infrared excess from dust debris has a short duration compared to the characteristic disk lifetime.  In the solar system, asteroids in the main belt have a mean inclination, $i$, of 7.9$^{\circ}$ (Binzel et al. 2000), and in white dwarf systems, it is unlikely that a newly tidally-disrupted asteroid has exactly the same orbital inclination as the pre-existing gaseous disk.  As a result, as the debris from
the asteroid orbits through the disk, it is eroded by sputtering.  

Let ${\Sigma}_{disk}$ (g cm$^{-2}$) denote the mean disk surface density:
\begin{equation}
{\Sigma}_{disk}\;=\;\frac{M_{disk}}{{\pi}D^{2}}
\end{equation}
If spherical grains of radius $a$ move with speed $v$ through a disk with number density, $n$, then the mass loss from sputtering is:
\begin{equation}
\frac{d}{dt}\left(\frac{4{\pi}{\rho}_{ast}\,a^{3}}{3\,m_{ast}}\right)\;=\;-y\,\left({\pi}\,a^{2}\,n\,v\right)
\end{equation}
where  $y$ is the sputtering yield.   If the grains and the disk have the same composition since they are both derived from asteroids, then
 each passsage through the disk erodes a layer of thickness, ${\Delta}a_{pass}$, where:
\begin{equation}
{\Delta}a_{pass}\;=\;\frac{y\,{\Sigma}_{disk}}{4\,{\rho}_{ast}\,\sin\,i}
\end{equation}
  With two passages per orbit, then in time $t$ with $t$$>>$ $t_{orbit}$, the minimum radius of a grain that survives, $a_{min}$  is: 
\begin{equation}
a_{min}\;=\;2{\Delta}a_{pass}\,\frac{t}{t_{orbit}}
\end{equation}

The infrared excess from a disk depends upon its optical depth which we now show rapidly becomes optically thin.  
Let ${\chi}$ denote
the opacity of the dust. Assuming that the grain optical cross section is given by its geometric
cross section and making the conservative assumption to assume that all the dust particles have size $a_{min}$, then:
\begin{equation}
{\chi}\;=\;\frac{3}{4\,{\rho}_{ast}\,a_{min}}
\end{equation}
If the mass of the asteroid is less than the mass of the disk, then 
the vertical optical depth through the disk, ${\tau}_{z}$, is bounded so that:
\begin{equation}
{\tau}_{z}\;{\leq}\;{\chi}\,{\Sigma}_{disk}
\end{equation}
Using equations (16) - (19), then:
\begin{equation}
{\tau}_{z}\;{\leq}\;\frac{3\,\sin\,i}{2\,y}\,\frac{t_{orbit}}{t}
\end{equation}
Light from the star of radius $R_{*}$, illuminates the disk at a slant angle of $R_{*}/D$, and the optical depth in the disk, ${\tau}_{disk}$, to light at this angle is:
\begin{equation}
{\tau}_{disk}\;{\approx}\;{\tau}_{z}\,\frac{D}{R{*}}\;=\;\frac{3\,D\,\sin\,i}{2\,R_{*}\,y}\,\frac{t_{orbit}}{t}
\end{equation}

From equation (21),  ${\tau}_{disk}$ diminishes with time as grains are destroyed.  We now
assume that the typical atom in the disk is silicon.  The orbital speed at the disk is near 500 km s$^{-1}$ and with an inclination angle of 7.9$^{\circ}$, the characteristic speed and kinetic energy of a silicon  atom impacting a grain are ${\sim}$70 km s$^{-1}$ and  ${\sim}$700 eV, respectively.  
Using the models of Tielens et al. (1994), we adopt $y$ ${\approx}$ 0.1.
For $D/R_{*}$ = 30, as representative of a disk within the tidal zone of the star, equation (21) shows that ${\tau}_{dist}$ becomes less than unity within ${\sim}$60 orbital times.  
 Since the orbital time of an asteroid might
be ${\sim}$10 years, this implies that there is appreciable infrared emission from the system for at most
${\sim}$600 years.  The true duration of an excess is probably much shorter since, for example, most incoming asteroids have masses appreciably smaller than that of the disk so that the bound given in equation (19) is conservative.  Also, we have
probably substantially overestimated the opacity of the dust debris by assuming all the particles
have the minimum size.  
\subsection{Duration of Dust Disks}

We propose that when a massive asteroid's orbit is perturbed so it is tidally destroyed, then a dust
disk results.  If analogous to Saturn's rings, this might only take ${\sim}$100 orbital times or ${\sim}$1000 years for the disk for form (Dones 1991).  The subsequent evolution of the dust disk depends upon
its viscosity which is highly uncertain.  If the disk were  composed purely of dust grains, then, like the rings of Saturn, the duration of the disk could be comparable to or longer than 10$^{8}$ yr.  However,
we anticipate that the dust disk is bombarded by orbitally-perturbed small asteroids and the resulting grain-grain collisions  produce gas in the disk.   For a mass distribution of asteroids given by equation (23) below, the median mass of a large asteroid that creates a dust disk is 2 $M_{disk}$.  A dust disk derived from the median-sized large asteroid  accumulates by further collisions of small asteroids, its own mass in gas in a time scale of 2 $t_{gas}$.   Furthermore, the asteroid may have had a substantial amount of
internal water which can be released into the gas phase as the asteroid is destroyed. Estimates for the percentage by mass of internal water in Ceres range between 0.17 and 0.27 (McCord \& Sotin 2005, Thomas et al. 2005).  If the viscous time scales inversely as the fraction of matter that is gaseous, then
in the absence of adding any additional gas from small asteroids, 
$t_{dust}$ would be between a factor of 4 and 6 longer than $t_{gas}$. Although extemely uncertain, we adopt on average that $t_{dust}$ ${\sim}$ 3 $t_{gas}$ or perhaps 1.5 ${\times}$ 10$^{5}$ yr.  

\section{MODEL ASSESSMENT}

We now  evaluate the model that asteroids are the main
source of contamination many externally-polluted white dwarfs.   In our ``standard model"  the mass of asteroid belt, $M_{belt}(t)$ is taken to equal
10$^{25}$ g at the onset of the white dwarf phase when $t$ = 0.  We assume that there is a  characteristic time, $t_{orbit}$,  to perturb an asteroid
into a tidally-disrupted orbit. Therefore:
\begin{equation}
M_{belt}\;=\;M_{belt}(0)\,e^{-t/t_{orbit}}
\end{equation}
This expression for an exponential decay of the asteroid belt is at best a rough approximation.
Wyatt et al. (2007) discuss models where the mass varies as $t^{-1}$ because of mutual collisions
while including orbital perturbations, more complex scaling models for the exponential decay may
apply (Dobrovolskis, Alvarellos \& Lissauer 2007).

A critical parameter in our analysis is $t_{orbit}$.  To-date,  simulations for the evolution of asteroid belts around white dwarfs have been somewhat limited,  although Duncan \& Lissauer (1998) and 
Debes \& Sigurdsson (2002) computed various scenarios.  For the future evolution of the solar system, Duncan \& Lissauer (1998) followed the orbits of 5 asteroids, and found that 1 was always unstable. In two out of their four simulations, a second asteroid, Pallas,  became unstable in 5 ${\times}$ 10$^{8}$ yr.   
To reproduce the data, we adopt values of $t_{orbit}$ near 1 Gyr.  
 However, a more simulations of  for asteroid belt evolution during the white dwarf phase are needed.
 
 We assume that the asteroid belt has a mass distribution, $n(M)\,dM$, given by a power law so that:
\begin{equation}
n(M)\,dM\;=\;A\,M^{-2}\,dM
\end{equation}
 where the constant $A$ is derived by the normalization condition. This power law variation of $M^{-2}$ is slightly steeper than the ``classical" estimate for the solar system that $n(M)$ varies as $M^{-1.83}$ (O'Brien \& Greenberg 2005), but it is computationally convenient and adequate for systems about which we know very little.   If $M_{belt}$ is the total mass of
 the asteroid belt with minimum mass, $M_{min}$ and maximum mass $M_{max}$, then
 \begin{equation}
A\;=\;M_{belt}\left(\ln\left[\frac{M_{max}}{M_{min}}\right]\right)^{-1}
\end{equation}
As in the solar system, we assume the largest asteroid has an appreciable fraction of the total mass of the entire system, and we adopt $M_{max}(t)$ = 0.2 $M_{belt}(t)$  Thus,  initially, the largest asteroid has a mass
of 2 ${\times}$ 10$^{24}$ g, somewhat larger than the mass of Ceres which is 9.4 ${\times}$ 10$^{23}$ g (Michalak 2000). We assume that $M_{min}$ is independent of time and equals 2 ${\times}$ 10$^{17}$ g, the mass of an asteroid with radius 3 km and density 2.1 g cm$^{-3}$. 
In this model, the total number of asteroids initially is 3 ${\times}$ 10$^{6}$.   With  $t_{orbit}$ = 1 Gyr, this would result in an asteroid being tidally disrupted every 300 yr.  
 
\subsection{Which contaminated white dwarfs have an infrared excess?}

In the context of our model, we now compute which white dwarfs may have an infrared excess.  This requires estimating $<{\dot M_{metal}}>$ which we take as:
 \begin{equation}
<{\dot M_{metal}}>\;=\;\frac{M_{0}}{t_{orbit}}\,e^{-t/t{orbit}}
\end{equation}
To compare this quantity with observables, we  estimate the white dwarf's  cooling time from its effective temperature.  We interpolate from the calculations
of Winget et al. (1987) and Hansen (2004) and take
\begin{equation}
t_{9}\;=\;0.38\,\left(\frac{12000}{T_{eff}}\right)^{3}
\end{equation}
where $t_{9}$ denotes the cooling time (Gyr).  This approach neglects the variation in stellar radii and masses.  
 If a white dwarf exhibits an accretion rate above $<{\dot M_{metal}}>$, then, according to our model,
 it is experiencing accretion from a single large asteroid and should possess an infrared excess.  
  We show in Figure 3 a plot of accretion rate vs. stellar effective temperature for externally-polluted hydrogen-rich white dwarfs where the accretion rate can be reliably estimated. In this plot, we also show the value of $<{\dot M_{metal}}>$ for   $M_{0}$ = 10$^{25}$ g and $t_{orbit}$ varying from 0.5 Gyr to 2.0 Gyr.  The model and the data are consistent for most but not all stars. The adopted estimated mass in the asteroid belts of 10$^{25}$ g is larger than in the solar system; this point is discussed below in ${\S6}$.  
 
 While the data points in Fig. 3 mostly obey the predicted relationship, there are exceptions. The most
 deviant point is for WD 1633+433 with ${\dot M}_{metal}$ = 4 ${\times}$ 10$^{8}$ g s$^{-1}$ and $T_{eff}$ = 6600 K (Koester \& Wilken 2006).  Perhaps this star has a particularly long-lived
 and  massive asteroid belt.  

\subsection{Relative numbers of contaminated white dwarfs with and without an infrared excess}
Any model to explain the contamination of white dwarfs should account for the relative numbers of
stars with and without an infrared excess.  We hypothesize in ${\S4}$  that an infrared excess occurs when
a tidally-disrupted asteroid is more massive than the ambient gaseous disk.
From equation (22), the number of such asteroids, $N(M>M_{disk})$, is:
 \begin{equation}
N(M>{M_{disk}})\;{\approx}\;\frac{A}{M_{disk}}
\end{equation}
The total rate at which these asteroids are perturbed into orbits where they are tidally-destroyed is
$N(M>M_{disk})/t_{orbit}$.  Therefore, if $t_{dust}$ is the duration of a dust disk,  the
fraction of time, $f_{IR}$, that the white dwarf exhibits an infrared excess is:
\begin{equation}
f_{IR}\;=\;N(M>M_{disk})\,\frac{t_{dust}}{t_{orbit}}
\end{equation}
Using equations from above,  this expression becomes:
\begin{equation}
f_{IR}\;=\;\left(\frac{t_{dust}}{t_{gas}}\right)\left(\ln\frac{M_{max}}{M_{min}}\right)^{-1}
\end{equation}
 Evaluation of expression (29) is insensitive to the exact values of $M_{max}$ and $M_{min}$.  From our estimates given above, we find that $f_{IR}$ = 0.062 $(t_{dust}/t_{disk})$.
In ${\S4.3}$, we argued
that $t_{dust}$ ${\sim}$ 3 $t_{disk}$.    Although extremely uncertain, the prediction from  equation (29) is therefore that $f_{IR}$ ${\sim}$ 0.19, consistent with the observational constraint determined by  
 Kilic \& Redfield (2007)  that at least 14\% of externally-polluted hydrogen-rich white-dwarfs possess an infrared excess.

\subsection{Composition of Accreted Material}

One motivation  for studying the composition of contaminated white dwarfs is that
it enables the measurement of the bulk elemental composition of extrasolar minor planets.  For example, the photosphere of the white dwarf  GD 362  was found to be  markedly deficient in volatiles such
as carbon and sodium relative to refractories such as calcium and iron (Zuckerman et al. 2007), a pattern  reminiscent of the inner solar
system where the Earth and asteroids also are very volatile deficient.  

Wolff et al. (2002) reported relative abundances of iron, magnesium, silicon, calcium and carbon in 10 helium-rich white dwarf.
Although in many cases the uncertainties are very large,   three of these stars exhibit a carbon abundance less than that of iron,  and therefore a value of $n$(C)/$n$(Fe) that is at least a factor of 10 smaller than solar.  These three stars have a composition similar to that of the Earth or chondrites; they do not appear to have accreted interstellar matter (Jura 2006).   
One of these three stars, GD 40, also displays an infrared excess (Jura et al. 2007b), consistent with the hypothesis that it is accreting chondritic material.  However, the other two stars do not display an infrared excess (Farihi et al. 2008b) which, however, as discussed below, does not preclude the asteroid
accretion scenario.

Desharnais et al. (2008) report FUSE data for 5 helium-rich white dwarfs.  Of these 5 stars,  GD 378  has $n$(C)/$n$(Fe) = 2.5, within the errors not grossly different from the solar value of 8.3 (Lodders 2003). There are mostly only upper limits  to the abundances  for GD 233, a second star in  this sample.    However, G270-124, GD 61 and GD 408 all have values of $n$(C)/$n$(Fe) less than 0.3. These three stars are therefore candidates for having accreted chondritic material.
  G270-124 does not display an infrared excess shortward of 8 ${\mu}$m (Mullally et al. 2007).  As far as we know, GD 61 has not observed for an infrared excess. 
There is a hint of an excess at 7.9 ${\mu}$m for GD 408 (Mullally et al. 2007), but the signal to noise
was not good enough for a definitive conclusion. 

Not all contaminated white dwarfs appear to have accreted chondritic material.  G238-44 has been found to display $n$(C)/$n$(Fe) near 2.5 (Dupuis et al. 2007);  this star does not
display any infrared excess (Mullally et al. 2007).  As with GD 378, the source of the external matter is unknown. 

There are at least two possible explanations for the absence of a near infrared excess around externally-polluted helium-rich white dwarfs.   First,  as discussed above, the circumstellar disk could be  largely gaseous.  Second, the disk may have dissipated before the contaminants have settled out of the star's atmosphere.  The settling time of heavy metals in the atmospheres of
these stars with effective temperatures between 15,000 K and 21,000 K ranges from ${\sim}$3 ${\times}$ 10$^{4}$ yr to ${\sim}$5 ${\times}$ 10$^{5}$ yr (Dupuis et al. 1993), comparable to or
longer than our very rough estimate of the duration of a dust disk of 1.5 ${\times}$ 10$^{5}$ yr.  

\subsection{Infrared excess from a largely gaseous disk?}

Above, we have adopted the view that a disk is either  essentially all dust or all 
gas.  However, 
it is possible that some dust survives in disks that are  bombarded by numerous small asteroids.
Consider the case where  the interior of the disk is gaseous yet the outer portion  remains dusty, and thus
 the disk distribution may have an especially large central hole.    All but one of the ten white dwarfs currently known to have an infrared excess has dust as warm as 1000 K (see Jura et al. 2007b). This innermost temperature may be determined by the location where grain sublimation becomes rapid.   However,   G166-58 (also known as LHS 3007 or WD 1456+298), a metal polluted white dwarf with a metal accretion rate of 3 ${\times}$ 10$^{7}$ g s$^{-1}$ (Koester \& Wilken 2006), was found by Farihi et al. (2008a) to have
 an excess at 5.7 ${\mu}$m and 7.9 ${\mu}$m, but not at shorter wavelengths.  Following Jura et al. (2007b), we compute a model with a flat disk to reproduce the data.  We adopt a stellar temperature of
 7390 K and a ratio of the star's radius (9.1 ${\times}$ 10$^{8}$ cm) to the star's distance from the Sun (${\approx}$30 pc), $R_{*}/D$, of
 1.0 ${\times}$ 10$^{-11}$.   We fit the infrared data with an inner disk temperature of 400 K (corresponding to a physical distance from the star of 30 $R_{*}$, Jura [2003]), an outer disk temperature of 300 K (corresponding to a physical distance from the star of 43 $R_{*}$), and $\cos i$ = 0.4.   This model disk lies well within the tidal zone which extends to ${\sim}$ 100 $R_{*}$ (Davidsson 1999). 
 The comparison between observations and the model is shown in Figure 4; the fit is good enough
 that this model is  a serious contender for explaining the data.  Thus, this system may be in an 
 intermediate phase of its disk evolution where the dust disk has an usually large hole because
 bombardment destroyed  dust  in the inner disk.  
 
 \subsection{Detectability of line emission from gaseous disks?}
 
 Gaensicke et al. (2006, 2007) report emission lines from two disks; most polluted white dwarfs  do not display
 emission lines even though we postulate the presence of a circumstellar reservoir from which they accrete.  We suggest that the systems with detected emission are unusual because the gas
 has an unusually high excitation temperature, $T_{ex}$.  If the star has effective temperature, $T_{*}$, and subtends solid angle,  ${\Omega}_{*}$, while the disk subtends solid angle, ${\Omega}_{disk}$, then a line is seen in emission if:
 \begin{equation}
B_{\nu}(T_{ex})\,{\Omega}_{disk}\;{\sim}\;B_{\nu}(T_{*})\,{\Omega}_{*}
\end{equation}
For SDSS 1228+1040, the analysis of Gaensicke et al. (2006) yields ${\Omega}_{disk}/{\Omega}_{*}$ = 1.2 ${\times}$ 10$^{4}$.  Since $T_{*}$ = 22020 K, then for the calcium lines near 8500 {\AA}, the above
criterion requires $T_{ex}$ $>$ 1800 K. This temperature is much greater than expected for a passively 
heated disk material lying at 100 stellar radii (see Jura 2003), and some additional heating of the 
gas must be occuring.  Gaensicke et al. (2006) show that the calcium lines are asymmetric and they explain this as the consequence of our viewing an eccentric disk.  Because the orbits should circularize by
mutual collisiions, it is plausible that the eccentricity is being driven by some external unseen planet.
If so, this material also would be heated.  Other disks may not be subjected to such extreme
heating, and their circumstellar gas would be too cool to be detected.

 \section{DISCUSSION}
There are a number of similarities between the asteroid belt in the solar system and the proposed asteroid belts around white dwarfs. 
We invoke asteroid belt masses of 10$^{25}$ g at the time the star enters its white dwarf phase, this  mass  is about a
factor of 6 larger than  the mass of the asteroid belt in the solar system.  However, if stars form with a Salpeter mass function, $n(M)\,dM$, that varies as $M^{-2.35}$ and if the lowest mass star that  currently is dying has a mass of 0.9 M$_{\odot}$, then the median main-sequence mass of a dying  star is ${\sim}$1.5 M$_{\odot}$.  There may be  a linear scaling between the  mass of a star's asteroid belt and its main-sequence mass. Furthermore, 1.5 M$_{\odot}$  stars spend
only ${\sim}$3 Gyr on the main-sequence before becoming white dwarfs, significantly less than the current age of the solar system of 4.6 Gyr.  Thus, when a 1.5 M$_{\odot}$ star becomes a white dwarf,  a simple linear extrapolation from the solar system suggests that it would possess 4 ${\times}$ 10$^{24}$ g, and thus approximately resemble the solar system.

A minimum mass of an extrasolar asteroid can be estimated from the mass of contaminants in the  atmosphere of a contaminated helium-rich white dwarf.
Since calculations for the mass in the outer mixing zone differ by as much as a factor of 10 (Dupuis et al. 1993, MacDonald et al. 1998), these estimates are uncertain.   At the moment, the most polluted known He-rich
white dwarf  is HS 2253+8023 (Wolff et al. 2002), and the inferred mass of its accreted asteroid is ${\sim}$10$^{24}$ g (Jura 2006), comparable to the mass of Ceres. 

The relative numbers of polluted white dwarfs with and without an infrared excess is explained by
a mass distribution of extrasolar asteroids that varies as $M^{-2}$, approximately what is observed in the solar system.  Thus extrasolar asteroid belts may possess a  size distribution similar to that of asteroids in the solar system.

Remarkably, in marked contrast to  the Sun,  carbon appears to be less abundant by number than  iron or silicon in
at least a few extrasolar asteroids.  Similarly, as measured from chondrites (Lodders 2003), asteroids in the solar system have a similarly low carbon to iron abundance ratio.    As argued previously (Jura 2006), determinations of the abundances in the atmospheres of white
dwarfs can serve both to evaluate  the model of interstellar accretion and, at least in some cases, to measure the composition
of extrasolar asteroids.  

\section{CONCLUSIONS}
 
 We suggest that  the 
 contamination of many externally-polluted white dwarfs without an infrared excess results from  tidal-disruptions of numerous small asteroids.   This scenario is consistent with the hypothesis that
 the masses,  size distributions and compositions of extrasolar  asteroids  are similar to those of the solar system's asteroids. 
 
  This work has been partly supported by NASA.  
        
 \begin{figure}
 \plotone{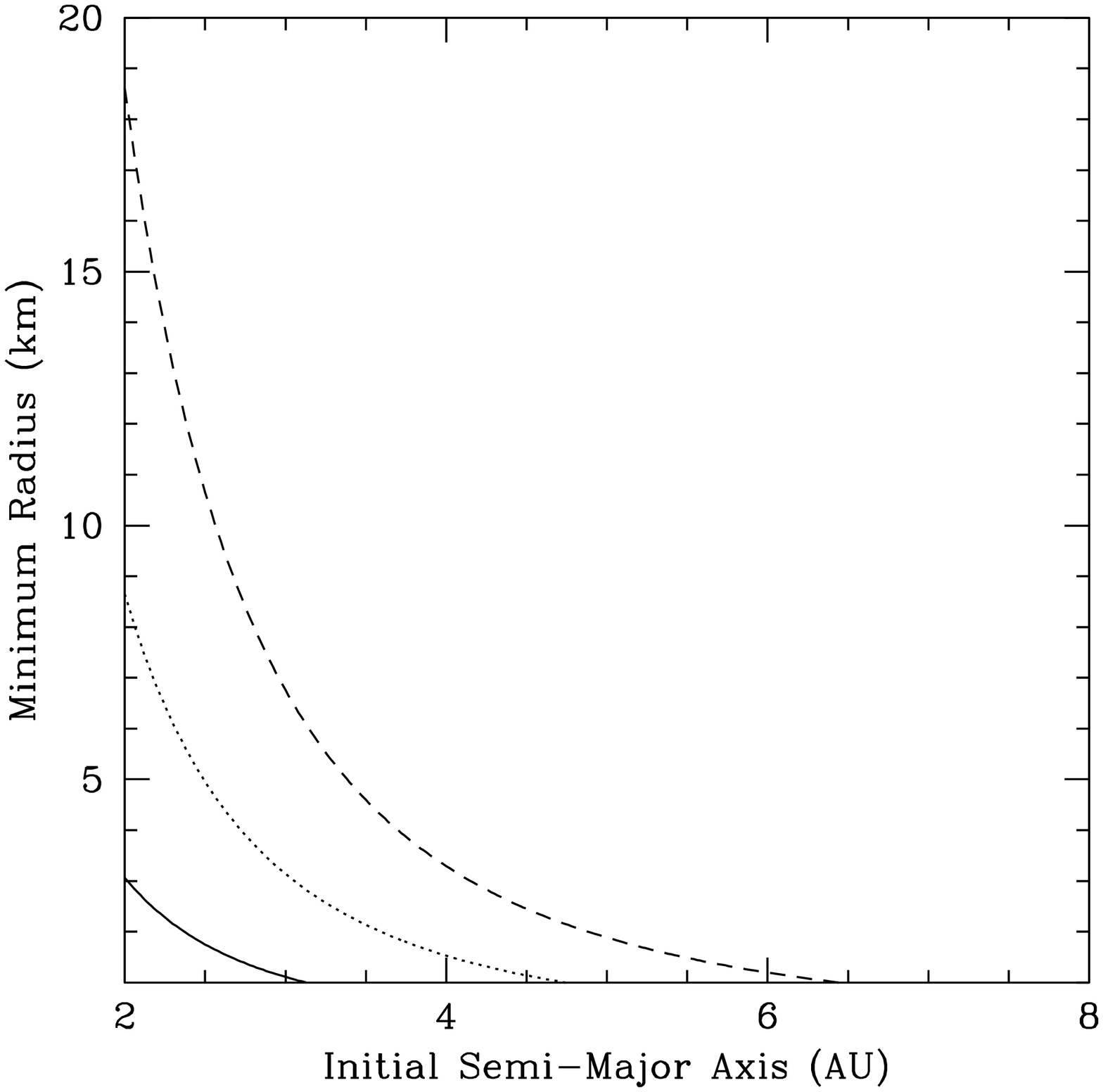}
 \caption{Minimum  surviving asteroid radius vs. initial distance from the star for stellar wind drag from equation (7). The solid, dotted and dashed lines are plotted for stars of initial masses of 1.5 M$_{\odot}$, 3.0 M$_{\odot}$ and 5.0 M$_{\odot}$, respectively.}
 \end{figure}
 \begin{figure}
 \plotone{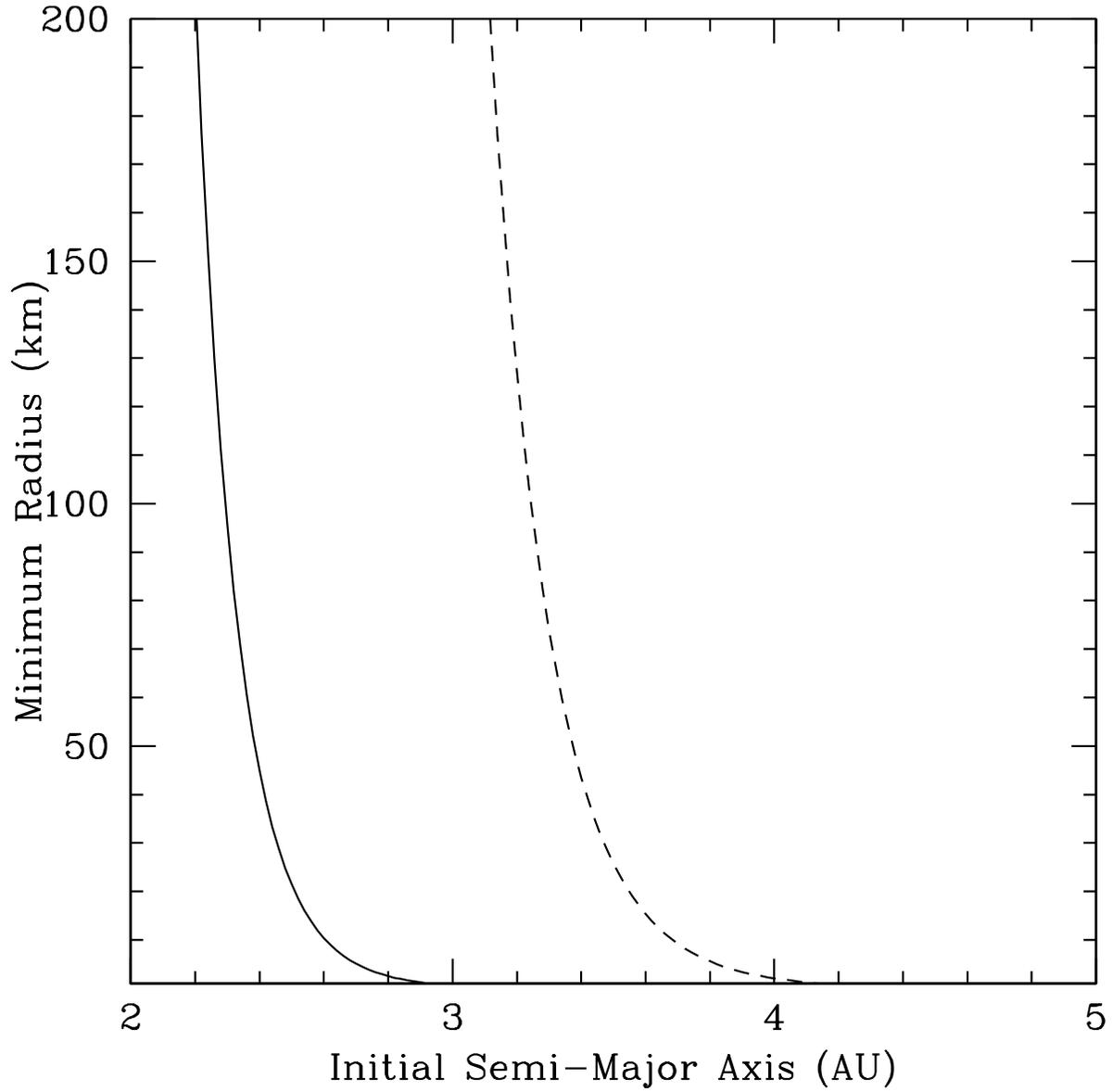}
 \caption{Minimum surviving asteroid radius vs. distance from the star for olivine sublimation (${\S}$3.2).  The solid line and dashed line are for stars with luminosities of 1.0 ${\times}$ 10$^{4}$ L$_{\odot}$ and 2.0 ${\times}$ 10$^{4}$ L$_{\odot}$, respectively.}
 \end{figure}
 \begin{figure}
 \plotone{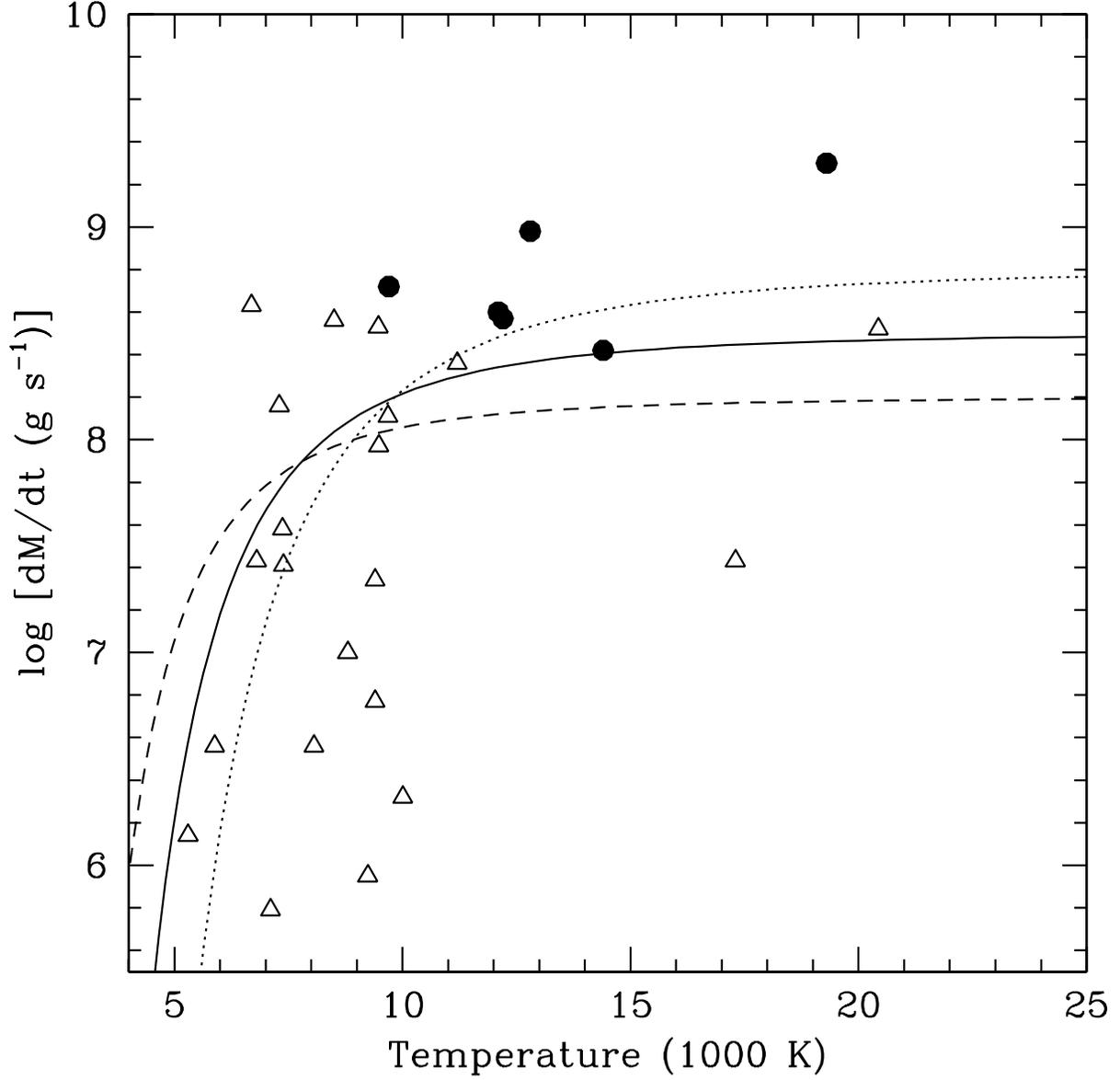}
 \caption{Metal accretion rates vs. stellar effective temperature.  The  filled circles represent stars with an infrared excess while the open triangles represent stars without an infrared excess. The data are taken from Farihi et al. (2008a) and  Jura et al. (2007b); the metal accretion rates are taken from Koester \& Wilken (2006) and scaled according to the procedure in Jura et al. (2007b).  The  lines represent
  $<{\dot M_{metal}}>$ as a function of the star's effective temperature from ${\S}$4.2.  We assume initial asteroid belt masses of 10$^{25}$ g and the dotted, solid and dashed lines represent values of $t_{depl}$  of 0.5 Gyr, 1.0 Gyr, and 2.0 Gyr, respectively. The solid circles are expected to lie above the curves while the open triangles are predicted to lie below the curves.}
 \end{figure}
 \begin{figure}
 \plotone{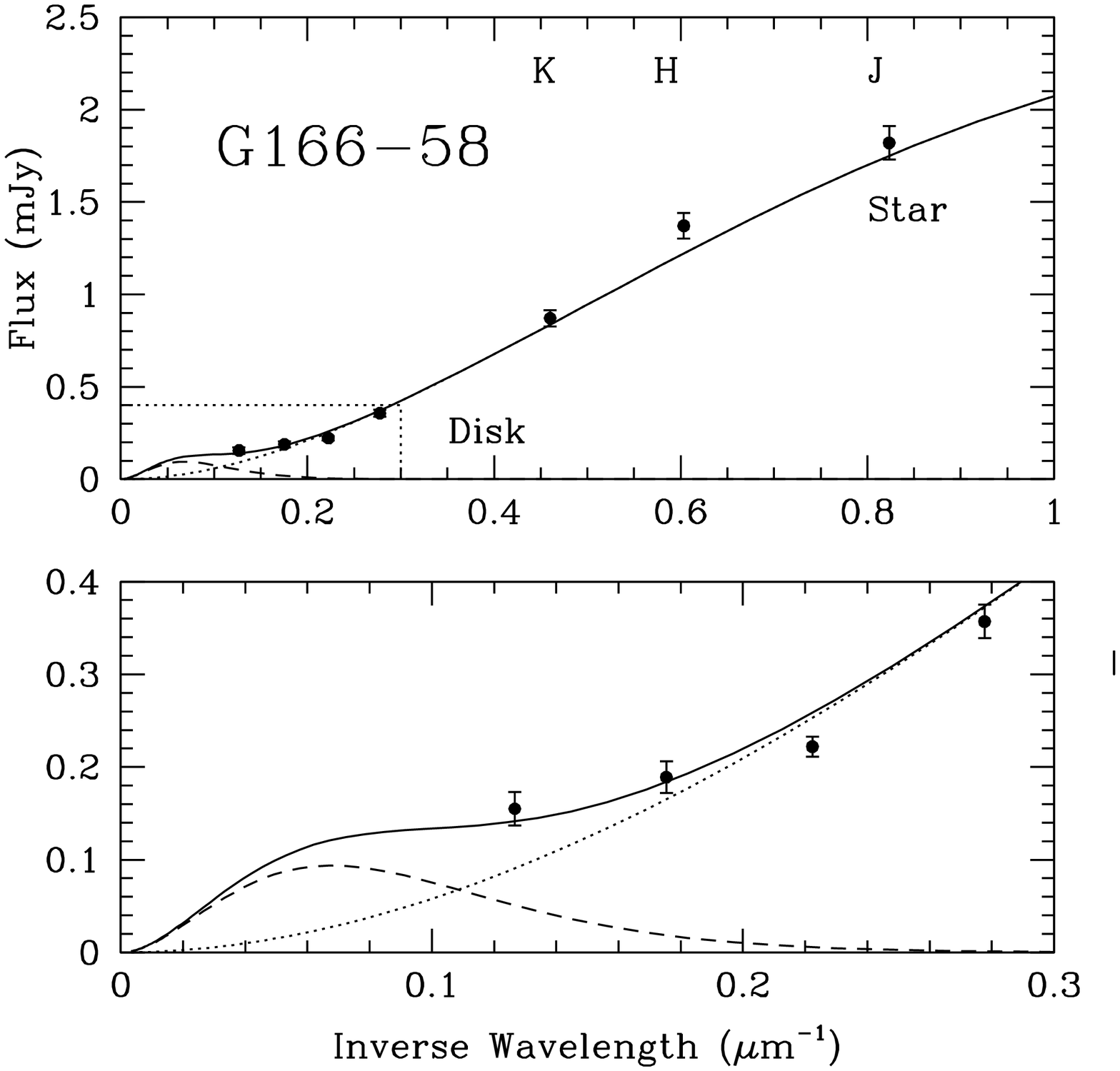}
 \caption{Comparison of data showing 1${\sigma}$ error bars (Farihi et al. 2008a) with the predicted stellar and disk emission for G166-58.  The model fluxes from the star, the disk and the total  are represented by the dotted, dashed and solid lines respectively.   For clarity, the lower panel re-displays  the dotted rectangle in the upper panel.}
 \end{figure}

\begin{thebibliography}{}
 \bibitem{becklin2005} Becklin,  E. E.,  Farihi, J., Jura, M., Song, I., Weinberger, A. J., \& Zuckerman, B. 2005, \apj, 632, L119
 \bibitem{binzel} Binzel, R. P., Hanner, M. S., \& Steel, D. I. 2000, in Allen's Astrophysical Quantitiies, ed. A. N. Cox (New York: Springer), 315
 \bibitem{Davidsson1999} Davidsson, B. J. R. 1999, Icarus, 142, 525
 \bibitem{debes2002} Debes, J. H., \& Sigurdsson, S. 2002, \apj, 572, 556
 \bibitem{desharnais2008} Desharnais, S., Wesemael, F., Chayer, P., Kruk, J. W., \& Saffer, R. A. 2008, \apj, 672, 540
 \bibitem{dones1991} Dones, L. 1991, Icarus, 92, 194
 \bibitem{dobrovolskis2007} Dobrovolskis, A. R., Alvarellos, J. L., \& Lissauer, J. J. 2007, Icarus, 188, 481
 \bibitem{dufour2007} Dufour, P. et al. 2007, \apj, 663, 1291
 \bibitem{duncan1998} Duncan, M. J. \& Lissauer, J. J. 1998, Icarus, 134, 303
 \bibitem{dupuis1993} Dupuis, J., Fontaine, G., Pelletier, C., \& Wesemael, F. 1993, \apjs, 84, 73
 \bibitem{dupuis2007} Dupuis, J., Bouabid, M.-P., Wesemael, F., \& Chayer, P. 2007, in 15th European
 Workshop on White Dwarfs, ASP Conference Series, v. 372, R. Napiwotzki \& M. R. Burleigh, eds. (San Francisco: Astronomical Society of the Pacific), 261
 \bibitem{farihi2008a} Farihi, J., Zuckerman, B., \& Becklin, E. E. 2008a, \apj, in press
 \bibitem{farihi2008b} Farihi, J., Zuckerman, B., Becklin, E. E., \& Jura, M. 2008b, \apj, in preparation
 \bibitem{friedrich2004} Friedrich, S., Jordan, S., \& Koester, D. 2004, \aap, 424, 665
 \bibitem{gaensicke2006} Gaensicke, B. T., Marsh, T. R., Southworth, J., \& Rebassa-Mansergas, A. 2006, Science, 314, 1908
 \bibitem{gaensicke2007} Gaensicke, B. T., Marsh, T. R., \& Southworth, J. 2007, \mnras, 380, L35 \bibitem{gonzalez2006} Gonzalez, G. 2006, \pasp, 118, 1494
 \bibitem{graham1990} Graham, J. R., Matthews, K., Neugebauer, G., \& Soifer, B. T. 1990, \apj, 357, 216
 \bibitem{hansen1999} Hansen, B. M. S. 2004, Phys. Rept., 399, 1
  \bibitem{jura2003} Jura, M. 2003, \apj, 584, L91 
  \bibitem{jura2006} Jura, M. 2006, \apj, 653, 613
  \bibitem{jura2007a} Jura, M., Farihi, J., Zuckerman, B., \& Becklin, E. E. 2007a, \aj, 133, 1927
  \bibitem{jura2007b} Jura, M., Farihi, J., \& Zuckerman, B. 2007b, \apj, 663, 1285
  \bibitem{jura1992} Jura, M., \& Kleinmann, S. G. 1992, \apjs, 79, 105
  \bibitem{keady1988} Keady, J. J., Hall, D. N. B., \& Ridgway, S. T. 1988, \apj, 326, 832
  \bibitem{kilic2007} Kilic, M., \& Redfield, S. 2007, \apj, 660, 641
  \bibitem{kilic2005} Kilic, M., von Hippel, T., Leggett, S. K., \& Winget, D. E. 2005, \apj, 632, L115
   \bibitem{kilic2006a} Kilic, M., von Hippel, T., Leggett, S. K., \& Winget, D. W. 2006a, \apj, 646, 474
 \bibitem{kilic2006b} Kilic, M., von Hippel, T., Mullally, F., Reach, W. T., Kuchner, M. J., Winget, D. W., \& Burrows, A. 2006b, \apj, 642, 1051
  \bibitem{kimura2002} Kimura, H., Mann, I., Biesecker, D. A., \& Jessberger, E. K. 2002, Icarus, 159, 529
  \bibitem{king2007} King, A. R., Pringle, J. E., \& Livio, M. 2007, \mnras, 376, 1740
   \bibitem{koester2006} Koester, D., \& Wilken, D.  2006, \aap, 453, 1051
   \bibitem{lodders2003} Lodders, K., 2003, \apj, 591, 1220
   \bibitem{macdonald1998} MacDonald, J., Hernanz, M., \& Jose, J. 1998, \mnras, 296, 523
   \bibitem{mccord2005} McCord, T. B., \& Sotin, C. 2005, J. Geophys. Res., 110, 5009
   \bibitem{michalak2000} Michalak, G. 2000, \aap, 360, 363
   \bibitem{mullally2007} Mullally, 
   F., Kilic, M., Reach, W. T., Kuchner, M. J., von Hippel, T., Burrows, A., \& Winget, D. E. 2007, \apjs, 171, 206
   \bibitem{obrien2005} O'Brien, D. P., \& Greenberg, R. 2005, Icarus, 178, 179
   \bibitem{paquette1986} Paquette, C., Pelletier, C., Fontaine, G., \& Michaud, G. 1986, \apjs, 61, 197
\bibitem{reach2005} Reach, W. T., Kuchner, M. J., von Hippel, T., Burrows, A., Mullaly, F., Kilic, M., \& Winget, D. E. 2005, \apj, 635, L161
\bibitem{reid1997} Reid, M. J., \& Menten, K. M. 1997, \apj, 476, 327
\bibitem{rybicki2001} Rybicki, K. R., \& Denis, C. 2001, Icarus, 151, 130
\bibitem{thomas2005} Thomas, P. C., Parker, J. Wm., McFadden, L. A., Russell, C. T., Stern, S. A., Sykes, M. V., \& Young, E. F. 2005, Nature, 437, 224
\bibitem{tielens1994} Tielens, A. G. G. M., McKee, C. F., Seab, C. G., \& Hollenbach, D. J. 1994, \apj, 431, 321
\bibitem{vonhippel2007} von Hippel, T., Kuchner, M. J., Kilic, M., Mullally, F., \& Reach, W. T.
 2007, \apj, 662, 544
 \bibitem{vonhippel2007b} von Hippel, T., \& Thompson, S. E. 2007, \apj, 661, 477
 \bibitem{weidemann2000} Weidemann, V. 2000, \aap, 363, 647
 \bibitem{winget1987} Winget, D. E., Hansen, C. J., Liebert, J., van Horn, H. M., Fontaine, G., Nather, R. E., Kepler, S. O., \& Lamb, D. Q. 1987, \apj, 315, L77
 \bibitem{wolff2002} Wolff, B., Koester, D., \& Liebert, J. 2002, \aap, 385, 995
 \bibitem{wyatt2007} Wyatt, M. C., Smith, R., Greaves, J. S., Beichman, C. A., Bryden, G., \& Lisse, C. 2007, \apj, 658, 569
 \bibitem{Zuckerman1980} Zuckerman, B. 1980, \araa, 18, 263
\bibitem{Zuckerman1987} Zuckerman, B., \& Becklin, E. E. 1987, Nature, 330, 138
 \bibitem{zuckerman2003} Zuckerman, B., Koester, D., Reid, I. N., \& Hunsch, M. 2003, \apj, 596, 477
 \bibitem{zuckerman2007} Zuckerman, B,, Koester, D., Melis, C., Hansen, B., \& Jura, M. 2007, \apj, 671,872
 \end{thebibliography}
\end{document}